\documentclass[acmsmall]{acmart}

\setcopyright{cc}
\setcctype{by}
\acmJournal{PACMMOD}
\acmYear{2025} \acmVolume{3} \acmNumber{3 (SIGMOD)} \acmArticle{164} \acmMonth{6} \acmPrice{}\acmDOI{10.1145/3725301}

\usepackage{amsthm}
\usepackage{amsmath}
\usepackage{amsfonts}
\usepackage{subcaption}
\usepackage{footmisc}
\usepackage{algorithm}
\usepackage{algpseudocode}
\usepackage{booktabs}
\usepackage{alltt}
\usepackage{enumitem}
\usepackage{wrapfig}
\usepackage{xcolor}

\usepackage[normalem]{ulem}

\newtheorem{example}{Example}
\newenvironment{example*}{Example \begin{centered}\itshape}{\end{centered}}

\newcommand{\Ione}[1]{{#1}}
\newcommand{\Itwo}[1]{{#1}}
\newcommand{\Ithree}[1]{{#1}}
\newcommand{\Ifour}[1]{{#1}}
\newcommand{\Ifive}[1]{{#1}}

\newcommand{\eat}[1]{}

\AtBeginDocument{%
  }

\begin{document}
\title{Galley: Modern Query Optimization for Sparse Tensor Programs}

\author{Kyle Deeds}
\email{kdeeds@cs.washington.edu}
\orcid{1234-5678-9012}
\affiliation{%
  \institution{University of Washington}
  \country{United States}
}

\author{Willow Ahrens}
\email{wahrens@mit.edu}
\orcid{1234-5678-9012}
\affiliation{%
    \institution{Massachusetts Institute of Technology}
    \country{United States}
}

\author{Magda Balazinska}
\email{magda@cs.washington.edu}
\orcid{1234-5678-9012}
\affiliation{%
  \institution{University of Washington}
  \country{United States}
}

\author{Dan Suciu}
\email{suciu@cs.washington.edu}
\orcid{1234-5678-9012}
\affiliation{%
  \institution{University of Washington}
  \country{United States}
}

\renewcommand{\shortauthors}{Kyle Deeds, Willow Ahrens, Magda Balazinska, \& Dan Suciu}

\begin{abstract}
The tensor programming abstraction is a foundational paradigm which allows users to write high performance programs via a high-level imperative interface. Recent work on {\em sparse tensor compilers} has extended this paradigm to sparse tensors (i.e., tensors where most entries are not explicitly represented). With these systems, users define the semantics of the program and the algorithmic decisions in a concise language that can be compiled to efficient low-level code. However, these systems still require users to make complex decisions about program structure and memory layouts to write efficient programs.  

This work presents \textit{Galley}, a system for declarative tensor programming that allows users to write efficient tensor programs without making complex algorithmic decisions. Galley is the first system to perform cost based lowering of sparse tensor algebra to the imperative language of sparse tensor compilers, and the first to optimize arbitrary operators beyond $\sum$ and $*$. First, it decomposes the input program into a sequence of aggregation steps through a novel extension of the FAQ framework. Second, Galley optimizes and converts each aggregation step to a concrete program, which is compiled and executed with a sparse tensor compiler. We show that Galley produces programs that are $1-300\times$ faster than competing methods for machine learning over joins and $5-20\times$ faster than a state-of-the-art relational database for subgraph counting workloads with a minimal optimization overhead.
\end{abstract}


\maketitle

\section{Introduction}

In recent years, the tensor programming (eq. array programming) model has become ubiquitous for high performance computing tasks. It has been applied to problems such as deep learning~\cite{DBLP:journals/corr/AbadiBCCDDDGIIK16,DBLP:conf/nips/PaszkeGMLBCKLGA19,DBLP:journals/nature/HarrisMWGVCWTBS20, DBLP:conf/sc/AndersonBDGMCHDBS90}, data cleaning~\cite{DBLP:conf/ssdbm/SunK023}, graph algorithms~\cite{DBLP:conf/ipps/SzarnyasB0KMMW21}, relational query processing~\cite{DBLP:journals/pvldb/HeNBSSPCCKI22,DBLP:journals/pvldb/AsadaFGGZGNBSI22,koutsoukos_tensors_2021}, and scientific computing~\cite{DBLP:journals/cphysics/WangICA22,stonebraker_scidb_2013,DBLP:journals/ascom/ModiLS21}, among others. While this approach was originally limited to dense arrays, data from many domains is fundamentally \emph{sparse} (i.e., most entries are a fill value like $0$), including graph data, one-hot encodings, relational data, 3D physics meshes, sparse neural networks, and others. However, modern tensor programming systems like NumPy, PyTorch, and sparse tensor compilers lack the advanced optimization capabilities of relational databases. Instead, users are forced to optimize their programs manually which is challenging and time consuming. In this work, we address this by introducing Galley, a system for declarative sparse tensor programming powered by advanced, cost-based program optimization.

Efficiently processing sparse tensors is challenging. Traditional tensor processing frameworks are collections of hand-optimized functions over \emph{dense} tensors~\cite{DBLP:journals/corr/AbadiBCCDDDGIIK16,DBLP:conf/nips/PaszkeGMLBCKLGA19,DBLP:journals/nature/HarrisMWGVCWTBS20, DBLP:conf/sc/AndersonBDGMCHDBS90}. To take advantage of sparsity, these frameworks need to provide implementations for every combination of input tensors' formats, resulting in spotty coverage for operations over sparse data \cite{ivanov2023sten}. Sparse tensor compilers (STCs) have been developed to automatically produce these implementations \cite{DBLP:journals/pacmpl/KjolstadKCLA17, bik_compiler_2022, DBLP:journals/corr/abs-2404-16730, shaikhha_functional_2022, hu_taichi_2019}. However, these compilers expose even more performance decisions than traditional frameworks, and they similarly lack automatic optimization capabilities.\footnote{Some systems separate declarative and imperative concerns with a scheduling language. However, the user still controls both aspects. For a more detailed description of the prototypical STC, we direct the reader to \cite{DBLP:journals/pacmpl/KjolstadKCLA17}.}

\begin{example}
    Consider Fig.~\ref{ex:stc_log_reg_program} which implements logistic regression inference in the language of Finch, an STC\cite{DBLP:journals/corr/abs-2404-16730}. Here, the user must choose the output format for the intermediate \texttt{R} (line 3). In this case, she chose a \texttt{Dense} rather than a \texttt{Sparse} format, which would be $\approx10\times$ slower. Then, the user chooses the loop order (lines 5-6). In this case, she chose $i$-then-$j$, which is asymptotically faster than $j$-then-$i$ because each out-of-order access to $X$ requires a full scan of the tensor. Finally, the user picks a merge algorithm for each loop that describes how to iterate through the non-zero indices (line 8). Here, $X$ is \texttt{iterated} through, and each non-zero $j$ is \texttt{looked up} in $\theta$. If she chose to \texttt{iterate} through $\theta$, each inner loop would scan the entire vector. Even for a simple kernel, these decisions represent a minefield of potential slowdowns.
\end{example}

\begin{figure}
\small
\begin{alltt}
 0. # Manually specified format for input tensors 
 1. FUNC log_regression(X::Dense(Sparse()), \(\theta\)::Dense())
 2.    # Manually defined intermediate format
 3.    R = Dense()
 4.    # Manually defined loop order
 5.    FOR i=_
 6.        FOR j=_
 7.            # Manually defined iteration algorithm
 8.            R[i] += X[i::iter,j::iter]*\(\theta\)[j::lookup] 
 9.        END
10.    END
11.    P = Dense()
12.    FOR i=_
13.        P[i] = \(\sigma\)(R[i::iter])
14.    END
15. END
\end{alltt}
\caption{Logistic regression implemented in the language of a sparse tensor compiler.}
\label{ex:stc_log_reg_program}
\end{figure}

In this paper, we propose \emph{Galley}, a system for declarative sparse tensor programming. Galley makes algorithmic decisions on the users' behalf, freeing them to focus on the high-level semantics of their program without sacrificing computational efficiency. It accepts input programs written in a declarative language, equivalent to the core of the NumPy API, and automatically produces an optimized STC implementation using the Finch compiler \cite{DBLP:journals/corr/abs-2404-16730}. To do so, it first restructures the program into a sequence of aggregation steps, minimizing total computation and materialization costs (Sec.~\ref{sec:logical_optimizer}). It then optimizes each step by selecting the loop order, the optimal formats for all intermediate tensors, and the merge algorithm for each loop (Sec.~\ref{sec:physical_optimizer}). These decisions are all guided by a system for estimating sparsity via statistics on the input tensors (Sec.~\ref{sec:sparsity}). \emph{Galley builds on fundamental principles from cost-based query optimization while developing new techniques that are specific to producing optimized code for sparse tensor compilers}.

Designing Galley required overcoming three key challenges. First, the high-level optimization requires a complex rewriting of the original program which must respect the algebraic properties of the program. \Ifive{We addressed this by introducing a novel extension of the FAQ framework that can handle \textit{arbitrary sparse tensor programs} \cite{DBLP:conf/pods/KhamisNR16}.} Second, STCs provide a vast design space for kernel implementations which makes the per-aggregate optimization challenging. Galley's physical optimizer searches through this space efficiently by separating concerns (loop order, output format, and intersection algorithm) and applying branch-and-bound optimization. Lastly, the computational cost of a sparse tensor program depends on the data distribution of the input data which complicates the optimization process. Galley produces these data-dependent cost estimates by leveraging the similarity of sparsity estimation and relational cardinality estimation. \Itwo{By overcoming these challenges, we have attempted to design Galley for a broad set of use cases ranging from sparse ML to graph algorithms and scientific simulations. To this end, we have incorporated Galley into the PyData/Sparse library which implements the full NumPy API for sparse arrays \cite{abbasi2018sparse,DBLP:journals/nature/HarrisMWGVCWTBS20}.}

\begin{example}
Let $A$, $B$, and $C$ be sparse matrices, and suppose that you want to compute the matrix chain $ABC$.  Because they do not consider the sparsity of the inputs, traditional systems will always perform this in the order $(AB)C$ where the intermediate, $AB$, is stored as a sparse matrix. When given this problem, Galley will optimize at runtime for the input's sparsities. This allows it to consider plans that are only efficient for specific inputs. For example, it may: 1)  re-order the operations to perform $BC$ before multiplying with $A$  2) store the intermediate as a dense matrix 3) transpose $B$ to iterate over the shared dimension first. In Sec. \ref{sec:experiments}, we show that this can provide a $\textbf{10x}$ speedup over state-of-the-art tensor frameworks for this example.
\end{example}

\noindent{\bf Contributions} We claim the following contributions:
\begin{itemize}[left=0pt]
    \item We present Galley, a system for declarative sparse tensor programming (Sec.\ref{sec:overview}).  Galley is the first system to perform cost based lowering of sparse tensor algebra to the imperative language of sparse tensor compilers, and the first to optimize arbitrary operators beyond $\sum$ and $*$. 
    
    \item Galley supports a \textit{highly expressive language} for sparse tensor algebra with arbitrary algebraic operators, aggregates within expressions, and multiple outputs (Sec.\ref{sec:overview}). 
    
    \item Galley performs \textit{cost-based logical optimization} with a novel extension of the variable elimination framework to handle arbitrary aggregations and pointwise operators (Sec.\ref{sec:logical_optimizer}). Galley performs \textit{cost-based physical optimization} to determine loop orders, tensor formats, and merge algorithms (Sec.\ref{sec:physical_optimizer}).
    
    \item We propose a \textit{minimal interface for sparsity estimation} to guide optimizations and implement two estimators (Sec.\ref{sec:sparsity}). 

    \item We evaluate Galley and show that it is $\textbf{1-300x}$ faster than hand-optimized kernels for mixed dense-sparse workloads and $\textbf{.25-100x}$ faster than a SOTA database for highly sparse workloads (Sec.\ref{sec:experiments}).

    \item We have implemented Galley as part of the PyData/Sparse sparse array project and the Finch tensor compiler\cite{Pydata, DBLP:journals/corr/abs-2404-16730}.
\end{itemize}

\section{Background}
\subsection{Tensor Index Notation}
\label{subsec:tensor_index_notation}
Input to Galley is written in an extended version of Einstein Summation (Einsum) notation that we call \emph{tensor index notation}\cite{albert1916foundation}. Traditional Einsum notation permits a single summation wrapped around a multiplication. For instance, you can describe triangle counting in a graph with adjacency matrix $E_{ij}$ using the following statement:
\begin{align*}
    t = \sum_{ijk}E_{ij}E_{jk}E_{ik}
\end{align*}
To capture the diverse workloads of tensor programming, we additionally allow the use of arbitrary functions for both aggregates and pointwise operations, nesting aggregates and pointwise operations, and defining multiple outputs. For example, a user could perform logistic regression to predict entities that might be laundering money. Then, they could filter this set based on whether the entities occur in a triangle in the transactions graph. This is represented by $\max_{jk}(E_{ij}E_{jk}E_{ik})$, which is $1$ if $i$ occurs in at least one triangle and $0$. This can be written in tensor index notation as:
\begin{align*}
    L_i &= \sigma(\sum_jX_{ij}\theta_j) > .5\\
    V_i &= L_i\cdot\max_{jk}(E_{ij}E_{jk}E_{ik})
\end{align*}

Tensor compilers like Halide, TACO, and Finch each build off of similar core notations, adding additional structures like FOR-loops to let users specify algorithmic choices \cite{DBLP:conf/pldi/Ragan-KelleyBAPDA13,DBLP:journals/pacmpl/KjolstadKCLA17,DBLP:journals/corr/abs-2404-16730}. Crucially, the vast majority of operations in array programming frameworks like NumPy can be expressed as operations in tensor index notation. Therefore, though we focus here on this notation, traditional tensor workflows can be captured and optimized in this framework.



\begin{figure*}[t]
    \centering
    \includegraphics[width=\linewidth]{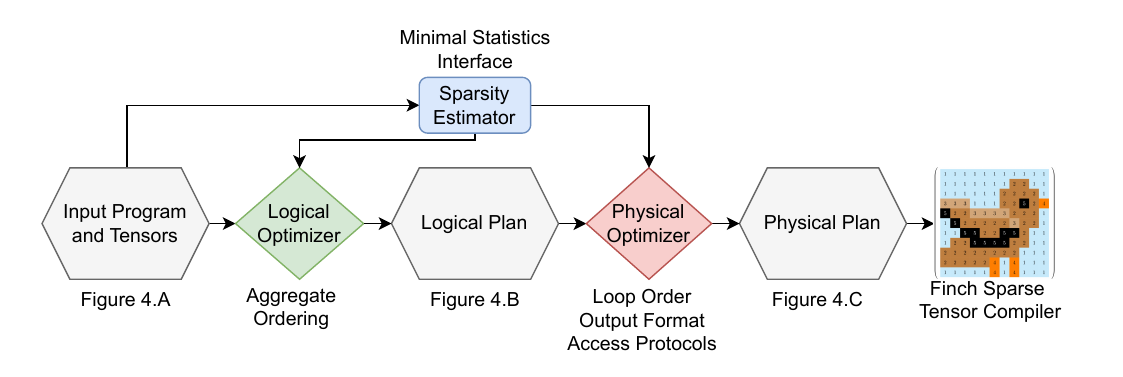}
    \caption{Galley overview.}
    \label{fig:overview}
\end{figure*}
\subsection{Sparse Tensor Compilers}
\label{subsec:sparse_tensor_compilers}
Over the last decade, compiler researchers have developed a series of sparse tensor compilers and shown that they produce highly efficient code for sparse tensor computations\cite{DBLP:journals/pacmpl/KjolstadKCLA17,DBLP:journals/corr/abs-2404-16730}. We use this work as our execution engine, so we briefly explain its important concepts below.
\begin{figure}[h]
    \centering
    \includegraphics[width=.9\linewidth]{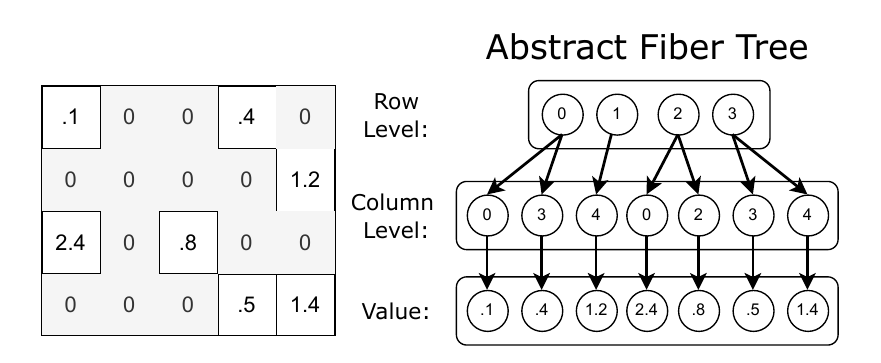}
    \caption{Fibertree format abstraction.}
    \label{fig:fiber_tree}
\end{figure}
\paragraph{Tensor Formats}
There are many different ways to represent sparse tensors, and the optimal approach depends on the data distribution and the workload. \Ifive{Work in this space has converged on the {\em fibertree} abstraction for describing the space of formats \cite{DBLP:journals/pacmpl/KjolstadKCLA17, sze_efficient_2017}.} In this formalism, a tensor format is a nested data structure resembling the one in Fig.~\ref{fig:fiber_tree}. Each layer stores the non-fill (e.g., non-zero) indices in a particular dimension, conditioned on earlier dimensions, and pointers to the next dimension's non-fill indices. These layers can be represented in any format that enables iteration and lookup. 

In this work, we consider sorted lists, hash tables, bytemaps, and dense vectors, which each perform differently in terms of iteration, lookup, and memory footprint. For example, the compressed sparse row (CSR) is a common format for sparse matrices. It stores the row dimension as a dense vector, where each entry points to the set of non-zero columns for that particular row. This set of non-zero columns is then stored in a sorted list, i.e., in a compressed sparse format. Importantly, this abstraction requires tensors to be accessed in the order in which they are stored (e.g., row-then-column in the case of CSR), which restricts the set of valid loop orders, as we describe next.

\begin{figure}[t]
    \raggedright    
    \begin{alltt}
\textbf{A.\,\,\,Input Program}
Plan := Query...    Query := (Name, Expr)
Agg := (Op, Idx..., Expr)   Map := (Op, Expr...)
Expr := Agg | Map | Input | Alias
Input := Tensor[Idx...]  Alias := Name[Idx...]
\textbf{B.\,\,\,Logical Plan}
Plan := Query...   Query := (Name, Agg)
Agg := (Op, Idx..., Expr)   Map := (Op, Expr...)
Expr := Map | Input | Alias
Input := Tensor[Idx...]  Alias := Name[Idx...]
\textbf{C.\,\,\,Physical Plan}
Plan := Query...   Query := (Name, Mat, Idx...)
Mat := (Format..., Idx..., Agg)
Agg := (Op, Idx..., Expr)   Map := (Op, Expr...)   
Expr := Map | Input | Alias   
Input := Tensor[PIdx...]  Alias := Name[PIdx...]
PIdx := Idx::Protocol
    \end{alltt}
    \caption{Query plan dialects.}
    \label{grammar:query_plans}
\end{figure}
\paragraph{Loop Execution Model}
The input to a Sparse Tensor Compiler is a high-level domain specific language (DSL); it consists of for-loops, in-place aggregates (e.g., $+=$), and arithmetic over indexed tensors (e.g., $A[i,j]*B[j,k]$). Crucially, the for-loops in these expressions are not executed in a dense manner. Instead, these compilers analyze the input formats and the algebraic properties of the expression to determine which index combinations will produce non-fill entries. In Fig.~\ref{ex:stc_log_reg_program}, because $0$ is the annihilator of multiplication (i.e., $x*0=0$), only the values of $i$ that map to non-zero entries in $X$ {\em and} $\theta$ are processed. All other index values will return a zero. So, the outer loop is compiled to an iteration over the intersection of the non-zero $i$ indices in $X$ and $\theta$; Fig.~\ref{fig:fiber_tree} shows how this is simply co-iteration over the top levels of their formats. The inner loop then iterates over the $j$ indices that are non-zero in $X[i,\_]$, i.e., the non-zero columns that occur in each row. 

\paragraph{Merge Algorithms}
Once the compiler has determined which tensors' non-zero indices must be merged to iterate over a particular index, it can apply several algorithms. All formats enable both ordered iteration and lookup operations; therefore, one algorithm iterates through the indices of all inputs, similar to a merge join, which is highly efficient per operation. However, this algorithm is linear in the total size of all inputs even if one is much smaller than the others. Another method is to iterate through a single input's level and lookup that index in the others. In this work, we take the latter approach, as described in Sec. \ref{subsec:merge_algorithms}. \Ifive{We refer to the mode of an individual tensor (such as ``iterate'' or ``lookup'') as an \emph{access protocol} and the overall strategy as a \emph{merge algorithm} \cite{ahrens_looplets_2023}.}

\section{Galley Overview}
\label{sec:overview}
We now provide a high-level view of Galley. We show how it transforms an input program to a logical plan then to a physical plan that is executed by an STC, as illustrated in Fig. ~\ref{fig:overview}. These steps are each represented by a dialect of our query plan language, whose grammar is defined in Fig.~\ref{grammar:query_plans}. In the following discussion, we use this grammar as a guide to show how our example program, i.e., logistic regression, would be transformed through these steps.

\subsection{Input Program Space}
The input program dialect is equivalent to the tensor index notation defined in Sec.~\ref{subsec:tensor_index_notation}. Pointwise functions such as  $A_{ij}*B_{jk}$ are represented with \texttt{Map}. Aggregates such as $\sum_i$ are denoted by \texttt{Agg}. Each assignment is a \texttt{Query}, and previous assignments are referenced with an \texttt{Alias}. \Ifive{Crucially, the \texttt{Op} terminal used in both \texttt{Map} and \texttt{Agg} can be any user defined function (e.g. $f(x,y) = sin(1+x*y)$) as long as it accepts the correct number of arguments (i.e. the number of expressions in the \texttt{Map} and two arguments in \texttt{Agg}). Galley takes advantage of properties of these functions during optimization, specifically distributivity, commutativity, associativity, identity, idempotency, and the existence of an annihilator. Further, users can declare these properties to Galley at runtime. This extensibility is a benefit of Galley’s formal framework. Lastly, \texttt{Idx}s are named symbols (e.g. $i$, $j$), and \texttt{Tensor}s are memory-resident input tensors.} Our logistic regression example from Fig. \ref{ex:stc_log_reg_program} is defined in this dialect as 
\begin{alltt}
Query(P, Map(\(\sigma\), Agg(+, j, Map(*, X[i,j], \(\theta\)[j]))))
\end{alltt}

Note that this notation is compatible with array APIs like Numpy that do not have named indices. Operations like 'matmul' can be automatically mapped into this language by generating index names for inputs on the fly and renaming whenever operations imply equality between indices.

\subsection{Logical Plan}
\label{subsec:logical_plan}
The first task in our optimization pipeline, handled by the logical optimizer, breaks down the input program into a sequence of simple aggregates. This is enforced by converting the input program (~\ref{grammar:query_plans}.A) to a logical plan (~\ref{grammar:query_plans}.B). This dialect is a restriction of the input dialect, where each query contains a single aggregate statement that wraps an arbitrary combination of \texttt{Map}, \texttt{Input}, and \texttt{Alias} statements. Intuitively, each logical query corresponds to a single STC kernel that produces a single intermediate tensor, but it does not specify details like loop orders and output formats. To perform this conversion soundly, each input query must correspond to a logical query, which produces a semantically equivalent output. To do this efficiently, 
Galley must minimize the total cost of all queries in the logical plan. 

Our logistic regression program above is not a valid logical plan because the outer expression is a pointwise function not an aggregate. However, it can be translated into the following logical plan 
\begin{alltt}
Query(R, Agg(+, j, Map(*, X[i,j], \(\theta\)[j])))
Query(P, Agg(no-op, Map(\(\sigma\), R[i])))
\end{alltt}
In this plan, the first query isolates the sum over the $j$ index, while the second query performs the remaining sigmoid operation on the result. Note that the latter query uses a no-op aggregate to represent an element-wise operation while conforming to the logical dialect.

\subsection{Physical Plan}
\label{subsec:physical_plan}
Given the logical plan, Galley's physical optimizer determines the implementation details needed to convert each logical query to an STC kernel. Specifically, it defines the loop order of each compiled kernel, the format of each output, and the merge algorithm for each index. As above, this is expressed by converting the logical plan to a physical plan described in the most constrained dialect. To avoid out-of-order accesses, we require that the index order of inputs and aliases are concordant with the loop order, so the physical optimizer may insert additional queries to transpose inputs. Therefore, each logical query corresponds to \emph{one or more} physical queries.

Using this language, we can precisely express the program from Fig.~\ref{ex:stc_log_reg_program} as follows, where \texttt{it} means iterate and \texttt{lu} means lookup.
\begin{alltt}
Query(R,Mat(dense,i,Agg(+,j,Map(*, X[i::it,j::it], 
                                            \(\theta\)[j::lu]))), i, j)
Query(P,Mat(dense,i, Map(\(\sigma\), P1[i::it])), i)
\end{alltt}
The first query computes the sum by iterating over the valid i indices for \texttt{X}, iterating over the j indices in the intersection of \texttt{X[i,\_]} and $\theta$, and materializing (hence \texttt{Mat}) their product in a dense vector over the i indices. The second query runs over this output and applies the sigmoid function, returning the result as a dense vector.

\subsection{Execution}
\label{subsec:execution}
Once Galley has generated a physical plan, the execution is very simple. For each physical query, it first translates the expression into an STC kernel definition and calls the STC to compile it. Then, Galley injects the tensors referenced by inputs and aliases and executes the kernel, storing the resulting tensor in a dictionary by name. After all queries have been computed, it returns the tensors requested in the input program by looking them up in this dictionary.

\section{Logical Optimizer}
\label{sec:logical_optimizer}
Given the plan dialects above, we now describe the logical optimizer, which receives an input program (Dialect ~\ref{grammar:query_plans}.A) and outputs a semantically equivalent {\em logical plan} (Dialect ~\ref{grammar:query_plans}.B). Specifically, the logical optimizer converts each query in the input program to a sequence of logical queries, where the last query produces the same output as the input query. There are many valid plans, and the optimizer searches this space to identify the cheapest one. We now briefly define "cheapest" in this context before outlining the complex space of logical plans that are considered. Finally, we explain the algorithms that we use to perform this search.

\subsection{Normalization \& Pointwise Distributivity}
The first step in logical optimization is to normalize the input program with a few simple rules that we apply exhaustively: (1) merge nested \texttt{Map} operators, (2) merge nested \texttt{Agg} operators, (3) lift \texttt{Agg} operators above \texttt{Map} operators, when possible, and (4) rename indices to ensure uniqueness. Applying these rules compresses the input program and makes our reasoning simpler in later steps by ensuring that operator boundaries are semantically meaningful.

Next, we consider whether to distribute pointwise expressions. Doing so may or may not yield a better plan because it both makes operations more sparse and produces larger expressions.
\begin{example}
    Consider the following expression which computes the loss function for the alternating least squares (ALS) algorithm and its distributed form:
\begin{align*}
    \sum_{ij}(X_{ij}-U_iV_j)^2 = \sum_{ij}X_{ij}^2 - 2\sum_{ij}X_{ij}U_iV_j + \sum_{i}U_i^2\sum_jV_j^2
\end{align*}
If all inputs are dense, the non-distributed form is more efficient because it results in fewer terms and has the same computational cost per term. However, if $X_{ij}$ is sparse and $U_i,V_j$ are dense, then the distributed form is more efficient because all terms can be computed in time linear w.r.t. the sparsity of $X_{ij}$. Note that the squaring operation here is a pointwise function, not a matrix multiplication.
\end{example}
To take advantage of this potentially asymptotic performance improvement, Galley performs a greedy search for the optimally distributed expression. At each step, it considers all single applications of distributivity in the expression. It then runs variable elimination for each (described later in this section) and computes the cost an optimal logical plan. If applying distributivity improved on the cost of the original expression, it continues. If not, it returns the optimal logical plan discovered so far. Lastly, we additionally consider the expression derived from applying distributivity exhaustively.

\subsection{Cost Model}
Overall, Galley's logical optimizer attempts to minimize the time required to execute the logical program. Because logical queries do not correspond to concrete implementations, our logical cost model aims to approximate this time without reference to the particular implementation that the physical optimizer will eventually decide on. This approximation considers two factors: (1) the number of non-fill entries in the output tensor and (2) the amount of computation (i.e., the number of FLOPs) needed to produce the output. The former corresponds to the size of the tensor represented by \texttt{Agg}, $nnz(\texttt{Agg})$, and the latter corresponds to the tensor size represented by the \texttt{MapExpr} within, $nnz(\texttt{MapExpr})$. We assume that the inputs are in memory;  hence, there is no cost for reading inputs from disk. We then perform a simple regression to associate each cost with a constant, and we add them to produce our overall cost, $c$, as follows:
\begin{align*}
    cost \approx a*nnz(\texttt{Agg}) + b*nnz(\texttt{MapExpr})
\end{align*}
To estimate $nnz(\texttt{Agg})$ and $nnz(\texttt{MapExpr})$, we use the sparsity estimation framework described in Sec. \ref{sec:sparsity}.

\subsection{Variable Elimination}
\Ifive{The core of our logical optimizer is an extension of the \textit{variable elimination} (VE) (eq. FAQ) framework \cite{dechter1999bucket, DBLP:conf/pods/KhamisNR16}. In its original context, this algorithm described a means of marginalization for probabilistic models by removing one variable at a time. When applied to our setting, it allows us to define the logical plan for an input query via an order on the indices being aggregated over, i.e., an {\em elimination order}. If we are given this order, we can construct a valid logical plan by iterating through the elimination order one index at a time in order to (1) identify the minimal sub-expression needed to aggregate over it, (2) create a new logical query representing the result of that sub-expression, and (3) replace it in the original query with an alias to the result. At the end of this process, the remaining query no longer requires any aggregation and therefore is itself a logical query.}

\begin{example}
Consider optimizing the following matrix chain multiplication:
\begin{align*}
    E_{im} = \sum_{jkl}A_{ij}B_{jk}C_{kl}D_{lm}
\end{align*}
The elimination order $jkl$ corresponds to a left-to-right multiplication strategy because eliminating $j$ from the expression first requires performing the matrix multiplication between $A$ and $B$. Eliminating $k$ then requires multiplying that intermediate result with $C$, and so on. Concretely, this produces the following sequence of logical queries: 
\begin{alltt}\small
    Query(I1, Agg(+, j, Map(*, A[i,j], B[j,k])))
    Query(I2, Agg(+, k, Map(*, I1[i,k], C[k,l])))
    Query(E, Agg(+, l, Map(*, I2[i,l], D[l,m])))
\end{alltt}
Similarly, the elimination order $lkj$ corresponds to a right-to-left strategy, and the order $klj$ to a middle-first strategy.
\end{example}
Unlike traditional VE for sum-product queries, we support complex trees of pointwise operators and aggregates. This makes identifying minimal sub-queries challenging since we must carefully examine the expression's algebraic properties. Given a strategy for this, the core problem of optimizing VE is to search the space of elimination orders for the most efficient one. In the worst case, this takes exponential time w.r.t. the number of indices being aggregated over. In the following sections, we describe how we identify minimal sub-queries and our search algorithm for finding the optimal elimination order.

\subsection{Identifying Minimal Sub-Expressions}
We now explain how to identify the minimal sub-expressions (MSEs) needed to eliminate an index. In sum-product expressions, the MSE is simply the product of the tensors that are indexed by it. For more complex input programs, we show that identifying MSEs corresponds to a careful traversal down the {\em annotated expression tree}, examining the algebraic properties of each operation to determine how to proceed.

\begin{figure}[h]
    \centering
    \includegraphics[width=.35\linewidth]{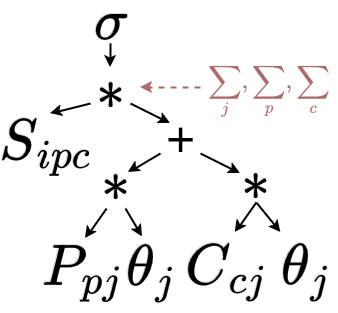}
    \caption{Annotated expression tree for logistic regression over joins $\sigma(\sum_{jpc}S_{ipc}(P_{pj}\theta_j + C_{cj}\theta_j))$}
    \label{fig:annotated_expression}
\end{figure}
\textbf{Annotated Expression Tree.} The annotated expression tree (AET) is constructed by examining the nested structure of \texttt{Agg}, \texttt{Map}, \texttt{Input}, and \texttt{Alias} nodes in the input query. To do this, Galley first removes all \texttt{Agg} nodes and annotates their inner expressions with \texttt{(Idx, Op)}. It then replaces all \texttt{Map} nodes with their operator to get the final tree, where every internal node is a pointwise function and every leaf is either an \texttt{Input} or an \texttt{Alias}.

\begin{example}
Fig.~\ref{fig:annotated_expression} shows the annotated expression tree for logistic regression where the input matrix is defined by a join-like expression $X_{ij}=S_{ipc}(P_{pj} + C_{cj})$. Further, Galley has pushed down $\theta_j$ into this expression. The sigmoid function is the outermost layer of the expression, so it appears at the top of the tree. The summations all occur just inside the sigmoid function, so they annotate the top multiplication operator. 
\end{example}

Given the AET, Galley identifies an index's MSEs by starting at the node where it is annotated and traversing downwards according to the algebraic properties of each internal node. We now describe the traversal rules for functions that are distributive, non-distributive, and commutative with respect to the aggregation operator.

\textbf{Distributive Functions.} When we reach a function that distributes over the aggregate (e.g., $*$ and $\sum$), we examine how many of the children, subtrees of the AET, contain the current index. If one child contains the index, we traverse down that child's branch, i.e., we factor the other children out of the aggregate. If multiple children contain the index, we wrap the sub-tree rooted at that node in the aggregate and return it as our MSE. If the function is commutative and associative, we only include the children that contain the index.

\textbf{Commutative, Identical Functions.} When the node's function is the same as the aggregate function and is commutative, we can push the aggregate down to each child independently. For example, we can transform the expression $\sum_{i} A_i + B_i$  into $\sum_i A_i + \sum_i B_i$. For all children that contain the index, we add the result of traversing down its branch to the list of MSEs and replace it with an alias to the result. If a child does not contain the index, then we need to account for the repeated application of the aggregate function. For example, $\sum_i B = N_i * B$ where $N_i$ is the size of the $i$ dimension.

\textbf{Blocking Functions}. A function that does not distribute or commute with our aggregate function is called a \textit{blocking function}. When we reach a blocking function in our traversal, we simply wrap it in our aggregate and return the sub-tree as an MSE. For example, the expression $\sum_j\sqrt{A_{ij}B_{jk}}$ cannot be rewritten as $\sqrt{\sum_jA_{ij}\sum_jB_{jk}}$ because $\sqrt{}$ is a blocking function.

\textbf{Discussion}. Galley builds upon and extends the theoretical FAQ framework for optimizing conjunctive queries with aggregation\cite{DBLP:conf/pods/KhamisNR16}. This framework explored the optimization of queries with the following form, where each $\bigoplus^{(i)}$ is either equal to or forms a semi-ring with $\otimes$:
\begin{align*}
    \bigoplus_{v_1}^{\,\,\quad(1)}\cdots\bigoplus^{\,\,\quad(k)}_{v_k} F^{1}_{V_1}\otimes\cdots\otimes F^k_{V_k}
\end{align*}
\Ifive{Similarly to Galley, the FAQ paper described the optimization problem as selecting an optimal elimination order over the aggregated variables.} Though this framework captures many important problems, it lacks the flexibility needed to support a general tensor processing system. Consider a slightly modified version of the SDDMM kernel:
\begin{align*}
    \sum_j A_{ik}(B_{ij} + C_{jk})
\end{align*}
This expression is not an FAQ query because it mixes addition and multiplication in the pointwise expression. Galley extends this framework to accommodate arbitrary pointwise expressions and placement of aggregates within expressions.

\subsection{Restricted Elimination Orders}
Depending on the program structure, the order in which indices can be eliminated might be restricted. This could be due to \textit{non-commutative aggregates} or \textit{aggregate placement}. The former is when an aggregate wraps another aggregate that it does not commute with. For example, given $\max_i\sum_j A_{ij}$, we must perform the summation before handling the maximum because $\max$ and $\sum$ do not commute. The latter is when an aggregate wraps another aggregate but cannot reach it via the traversal described above, e.g., $\sum_i\sqrt{\sum_j A_{ij}}$; in this case, the inner aggregate must be performed first. Collectively, these restrictions form a partial ordering on the index variables that must be respected when we enumerate elimination orders.

\subsection{Search Algorithms}
With the VE approach, we have simplified the complicated issue of high-level optimization to the discrete problem of choosing an optimal order on the aggregated index variables. We start by revisiting our example from Fig.~\ref{fig:annotated_expression}. The input query is the following,
\begin{alltt}
    Query(X, Map(\(\sigma\), 
                Agg(+,p,c,j,
                    Map(*,S[i,p,c], 
                        Map(+, 
                            Map(*, P[p,j], \(\theta\)[j]),
                            Map(*, C[c,j], \(\theta\)[j]))))))
\end{alltt}
The elimination order for this expression is an ordering of the indices $\{p, c, j\}$. Galley's logical optimizer searches through these possible orders to find the most efficient one. In this case, it would choose $[j,p,c]$, resulting in the following logical plan,
\begin{alltt}
    Query(A1, Agg(+, j, Map(*, P[p,j], \(\theta\)[j])))
    Query(A2, Agg(+, j, Map(*, C[c,j], \(\theta\)[j])))
    Query(A3, Agg(+, p, c, Map(*, S[i,p,c],
                                Map(+, A1[p], A2[c]))))
    Query(X, Map(\(\sigma\), A3[i]))
\end{alltt}
We now present two algorithms to search for that optimal order using the tools described above.

\textbf{Greedy.} The greedy approach chooses the cheapest index to aggregate at each step by finding the minimal sub-query for each index and computing its cost. The cheapest index's minimal sub-query is removed from the expression, appended to the logical plan, and replaced with an alias to the result. This continues until no aggregates remain in the expression.

\textbf{Branch-and-Bound.} The branch-and-bound approach computes the optimal variable order and occurs in two steps. The first step uses the greedy algorithm to produce an upper bound on the cost of the overall plan; the second performs a dynamic programming algorithm. In the dynamic programming step, the keys of the memo table are unordered sets of indices, and the values are tuples containing a partial elimination order, residual query, and cost. The algorithm initializes the table with the empty set and a cost of zero. At each step, it iterates through table entries and attempts to aggregate out another index. It then uses the cost bound from the first step to prune entries from the memo table whose cost exceeds the bound; doing so is valid because costs monotonically increase as more indices are added to the set. At the end of this step, the algorithm returns the index order associated with the full set of indices. 

\section{Physical Optimizer}
\label{sec:physical_optimizer}
Each query in the logical dialect roughly corresponds to a single loop nest and materialized intermediate. However, several decisions remain about {\em how} the kernel is computed, including: (1) the loop order over the indices,  (2) the format of the result, and (3) the merge algorithm for each index. The physical optimizer makes these decisions.

\subsection{Loop Order}
\label{subsec:loop_order}
The loop order determines that inputs are accessed. An good loop order prunes the iteration space due to early intersection of sparse inputs. Intuitively, this is similar to selecting a variable order for a worst-case optimal join algorithm. Galley's physical optimizer searches the space of loop orders to find one with the minimum cost, defined below.
\paragraph{Cost Model} The cost of a loop order is composed of each loop's number of iterations and the cost of transposing inputs to make them concordant with the loop order.
\begin{example} 
Consider matrix chain multiplication over three sparse matrices, $A$, $B$, and $C$, where 
\begin{align}
    D[il] = \sum_{jk} A[ij]*B[jk]*C[kl]
\end{align}
Suppose that $A$ has only a single non-zero entry and that $B$ and $C$ have 5 non-zero entries per column and per row. In this case, the loop order $ijkl$ is significantly more efficient than $lkji$. In the former, the first two loops, over $i$ and $j$, incur only a single iteration because they are bounded by the size of $A$. The third and fourth incur $5$ and $5^2$ iterations, respectively, because there are only $5$ non-zero $k$'s per $j$ in $B$ and $5$ non-zero $l$'s per $k$ in $C$. In the latter, the first two loops iterate over the full matrix $C$ despite most of those iterations not leading to useful computation.
\end{example}

Formally, let $Q$ be the pointwise expression in our kernel, and let $Q_{(i_1,\ldots,i_k)}$ be the restriction of that expression to just the index variables $i_1,\ldots,i_k$. Let $\mathbf{A}^{(i_1,\ldots,i_k)}$ be the input tensors that are not concordant with $i_1,\ldots,i_k$. Then, we can define the cost of a loop order as follows,
\begin{align*}
    cost(Q,(i_1,\ldots,i_k)) \approx \sum_{j=1}^knnz(Q^{(i_1,\ldots,i_j)}) + \sum_{A\in\mathbf{A}^{(i_1,\ldots,i_k)}}|A|
\end{align*}
In practice, we further refine this model to take into account the number and kind of tensor accesses at each loop.

\paragraph{Optimization Algorithm.} To optimize the loop order, we combine this cost model with a branch-and-bound, dynamic programming algorithm. In the first pass, the optimization algorithm selects the cheapest loop index at each step until reaching a full loop order. This produces an upper bound on the optimal execution cost, which the algorithm uses to prune loop orders in the second step. This step applies a dynamic programming algorithm. Taking inspiration from Selinger's algorithm for join ordering, each key in the DP table is a set of index variables and a set of inputs. The former represents the loops that have been iterated so far, and the latter represents a set of inputs that must be transposed.
 
\subsection{Intermediate Formats}
\label{subsec:intermediate_formats}
Once the loop order has been determined, the physical optimizer selects the optimal format for each query's output. First, Galley sets the order of the indices to be concordant with either the loop order of the kernel where it will be consumed or the order requested by the user. Then, it selects a format for each index (e.g., dense vector, hash table, etc.). Two factors affect this decision: (1) the kind of writes being performed (sequential vs random) and (2) the sparsity of the tensor at this index. The former is important because many formats (e.g., sorted list formats) only allow sequential construction. These formats can only be used if the output indices form a prefix of the loop order. 

When considering sparsity, Galley balances the fact that dense formats have better baseline efficiency, while sparse formats are asymptotically more efficient for highly sparse outputs. To describe this trade-off, we hand selected sparsity cutoffs between fully sparse, bytemap, and fully dense formats. To determine a particular output index's format, the physical optimizer  first determines the sparsity at this index level and uses our cutoffs to determine which category of formats to consider. Then, it checks whether sequential or random writes are being performed and selects the most efficient format that supports the write pattern.

\subsection{Merge Algorithms}
\label{subsec:merge_algorithms}
The final decision the physical optimizer makes concerns the algorithm it will use to perform each loop's intersection. While there are more complex strategies, we adopt instead a minimal approach and select a single input to iterate over for each loop. The physical optimizer then probes into the remaining inputs. It makes this selection by estimating the number of non-zero indices that each input has, conditioned on the indices in the outer loops. This resembles the approach taken in \cite{DBLP:journals/sigmod/WangWS24} for optimizing WCOJ.

\subsection{Common Sub-Expression Elimination}
Galley takes a straightforward approach to avoiding redundant computation. Once a physical plan has been generated, the right hand side of each physical query is canonicalized and hashed. When two physical queries result in the same hash, the latter query is removed from the plan and all references to it are replaced with a reference to the result of the former. This is helpful for caching small computations like transpositions, but it is also useful for reducing the overhead of applying distributivity which often results in duplicate sub-expressions.

\section{Sparsity Estimation}
\label{sec:sparsity} 
We now describe how Galley performs the sparsity estimation that guides our logical and physical optimizers. First, we explore the subtle correspondence between sparsity and cardinality estimation. We then present a minimal interface for sparsity estimation inspired by this correspondence, after which we examine two implementations of this framework, i.e., the uniform estimator and the chain bound.

\subsection{Sparsity and Cardinality Estimation} Sparsity estimation is highly related to cardinality estimation in databases. However, translating methods for the latter to the former requires analyzing the algebraic properties of our tensor programs. For example, let $A_{ij}$ and $B_{jk}$ be sparse matrices with a fill value of $0$, and let $R_A(I,J)$ and $R_B(J,K)$ be relations that store the indices of their non-zero entries. Assume we are performing the following, 
\begin{align*}
    C_{ijk} = A_{ij}B_{jk}
\end{align*}
In this case, the number of non-zero values in $C$ is precisely equal to the size of the conjunctive query
\begin{align*}
    nnz(C) = |R_A(I,J)\bowtie R_B(J,K)|
\end{align*}
The correspondence results from the fact that $0$ is the annihilator of multiplication (i.e., $x*0=0\forall x$), so any non-zero entry $ijk$ in the output must correspond to a non-zero $ij$ in $A$ \emph{and} a non-zero $jk$ in $B$. Consider the following instead:
\begin{align*}
    C_{ijk} = A_{ij} + B_{jk}
\end{align*}
In this case, a nonzero $ijk$ in the output can result from a non-zero $ij$ in $A$ \emph{or} a non-zero $jk$ in $B$. In traditional relational algebra, where relations are over infinite domains, this kind of disjunction would result in an infinite relation. However, tensors have finite dimensions, so we can introduce relations that represent the finite domains of each index, e.g., $D_i=\{1,...,n_i\}$. This lets us represent the index relation of the output as
\begin{align*}
    nnz(C) = |(R_A(I,J)\bowtie D_k(K))\cup (D_i(I)\bowtie R_B(J,K))|
\end{align*}
Finally, we can translate aggregations to the tensor setting as projection operations. Given the statement
\begin{align*}
    C_{ik} = \sum_j A_{ijk}
\end{align*}
we can express the non-zeros entries of $C$ as
\begin{align*}
    nnz(C) = |\pi_{I,K}(R_A(I,J,K))|
\end{align*}

\subsection{The Sparsity Statistics Interface}
We use our statistics interface to annotate every node of the AST with statistics. Surprisingly, to support sparsity estimation over arbitrary tensor algebra expressions, we only need a few core functions: (1) a constructor, which  produces statistics from a materialized tensor for \texttt{Input} and \texttt{Alias} nodes, (2-3) a function for (non) annihilating \texttt{Map} nodes (i.e., those whose children's fill values are the annihilator of its pointwise function), which merges the children's statistics, (4) a function for \texttt{Agg}, which adjusts the input's statistics to reflect an aggregation over some set of indices, and (5) an estimation procedure, which estimates the sparsity of a tensor based on its statistics.

\subsection{Supported Sparsity Estimators}

\subsubsection{Uniform Estimator}
The simplest statistic that can be kept about a tensor is the number of non-fill (e.g., non-zero) entries. The uniform estimator uses only this statistic and assumes these entries are uniformly distributed across the dimension space. This corresponds to System-R's cardinality estimator with the added assumption that the active domain is the whole dimension for each index~\cite{DBLP:conf/sigmod/SelingerACLP79}.

\textbf{Constructor.} This function simply counts the non-fill values in the tensor, $nnz(A)$, and notes the dimension sizes $n_{i_1},\ldots,n_{i_k}$.
\paragraph{Map (Annihilating).} To handle an annihilating pointwise operation, this function calculates the probability that a point in the output was non-fill in all inputs, then multiplies this by the dimension space of the output. For a set of inputs $A^{(1)}_{I_1}\ldots A^{(l)}_{I_l}$ and output $C_{I_C}$, where each $I_j$ is a set of indices, this probability is 
\begin{align*}
    nnz(C)\approx\left(\prod_{i\in I_C}n_i\right)\cdot\left(\prod_{j}\frac{nnz(A_j)}{\prod_{i\in I_j}n_i}\right)
\end{align*}

\textbf{Map (Non-Annihilating).} To handle an non-annihilating pointwise operation, this function calculates the probability that an entry in the output was \emph{fill} in all inputs. Then, it takes the compliment to get the probability that it was non-fill in all inputs and multiplies this by the output dimension space. Using the preceding notation:
\begin{align*}
    nnz(C)\approx\left(\prod_{i\in I_C}n_i\right)\cdot\left(1-\prod_{j}\left(1-\frac{nnz(A_j)}{\prod_{i\in I_j}n_i}\right)\right)
\end{align*}
\textbf{Aggregate.} Given an input tensor $A_{I}$ to aggregate over the indices $I'$, this function computes the probability that an output entry is non-fill by calculating the probability that at least one entry in the subspace of the input tensor was not fill:
\begin{align*}
    nnz(C)\approx\left(\prod_{i\in I\setminus I'}n_i\right)\cdot\left(1-\left(1-\frac{nnz(A_I)}{\prod_{i\in I}n_i}\right)^{\prod_{i\in I'}n_i}\right).
\end{align*}
\textbf{Estimate.} The estimation function simply returns the current tensor's stored cardinality statistic.
\begin{example}
    \Ifive{Suppose $A_{ij}$ and $B_{jk}$ are $100x100$ sparse matrices with $nnz(A)=1000$, $nnz(B)=200$, and we want to estimate $nnz(\sum_jA_{ij}B_{jk})$. We first compute $nnz(A_{ij}B_{jk})$ as $100^3 * \frac{1000}{100^2}*\frac{200}{100^2} = 2000$. The fractions are the probability that $i,j$ or $j,k$ was zero in $A$ or $B$, respectively. Next, we factor in the aggregation over $j$ to get $nnz(\sum_{j}A_{ij}B_{jk}) = 100^2 * (1-(1-\frac{nnz(A_{ij}B_{jk})}{100^3})^{100}) \approx 1800$. Here, the expression $(1-\frac{nnz(A_{ij}B_{jk})}{100^3})^{100}$ is the probability that all entries were zero for a particular $i,k$ pair.}
\end{example}
\subsubsection{Degree Statistics and the Chain Bound}
\label{subsubsec:chain_bound}
Galley stores degree statistics by default uses them to compute upper bounds on tensors' sparsities. A degree statistic, denoted as $D_A(X|Y)$, stores the maximum number of non-fill entries in the $X$ dimensions conditioned on the $Y$ dimensions for a tensor $A$. For example, given a matrix $A_{ij}$, $D_A(i|j)$ is the maximum number of non-fill entries per column, and $D_A(ij|\emptyset)$ is the total number of non-fill entries in the matrix. This approach follows work in cardinality bounding that  has been shown to produce efficient query plans in the relational setting \cite{DBLP:journals/pacmmod/DeedsSB23,DBLP:conf/cidr/HertzschuchHHL21,DBLP:conf/sigmod/CaiBS19}.

\textbf{Constructor.} This function first computes the boolean tensor representing the input's sparsity pattern. Then, to calculate each degree statistic, it sums over the $X$ dimensions and takes the maximum over the $Y$ dimensions. The set of degree statistics for a tensor $A_{I}$ is denoted $\mathcal{D}_{A_I}$.

\textbf{Map (Annihilating).} Annihilating map operations can only reduce the degree for any $X$, $Y$ pair. Therefore, every input's degree statistics are also valid for the output. If the inputs are $A^(1)_{I_1},...,A^(k)_{I_k}$, then the output's statistics are,
\begin{align*}
    \mathcal{D}_C = \bigcup_{j}\mathcal{D}_{A^{(j)}_{I_j}}
\end{align*}

\textbf{Map (Non-Annihilating).} In this case, Galley extends the degree constraints from each input to cover the full set of indices. Then, it computes degree statistics about the output, $C$, from the inputs $A^{(1)}_{I_1},\ldots,A^{(k)}_{I_k}$ by addition: 
\begin{align*}
    D_C(X|Y) = \sum_{j}D_{A^{(j)}}(X|Y)
\end{align*}

\textbf{Estimator.} This function calculates an upper bound (eq. performs sparsity estimation) using the breadth-first search approach described in \cite{DBLP:journals/sigmod/ChenHWSS23}. Intuitively, each set of indices forms a node in the graph, and each degree constraint is a weighted edge from $Y$ to $X$. Its search begins with the empty set; it then uses a breadth-first search to find the shortest weighted path to the full set of indices $I$. The product of the weights along this path bounds the number of non-zeros in the result.
\begin{example}
    \Ifive{Suppose $A_{ij}$ and $B_{jk}$ are $100x100$ sparse matrices with $D_A(ij|\emptyset)=1000, D_A(i|j)=10, D_B(jk|\emptyset) =200$, and we want to bound $nnz(\sum_jA_{ij}B_{jk})$. Because multiplication is an annihilating operation in this case, the degree constraints of $\sum_jA_{ij}B_{jk}$ are simply the union of the constraints for $A$ and $B$. To get a bound, we start by conditioning on the empty set and try to reach the output's index set, $i,k$, via the constraints, e.g. $D_B(jk|\emptyset) * D_A(i|j) = 2000$.}
\end{example}

\begin{figure*}
    \centering
    \includegraphics[width=\linewidth]{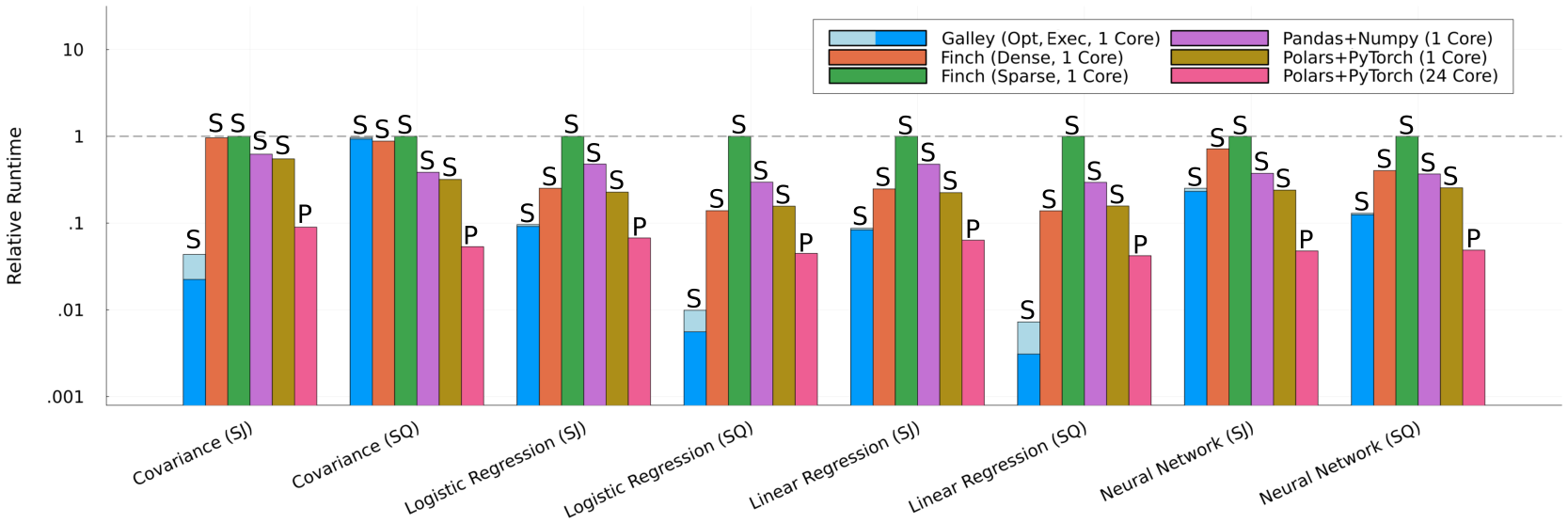}
    \caption{ML Inference Over Joins}
    \label{fig:ml_over_joins}
\end{figure*}

\begin{figure*}
\centering
\begin{subfigure}{.49\textwidth}
  \centering
  \includegraphics[width=\linewidth]{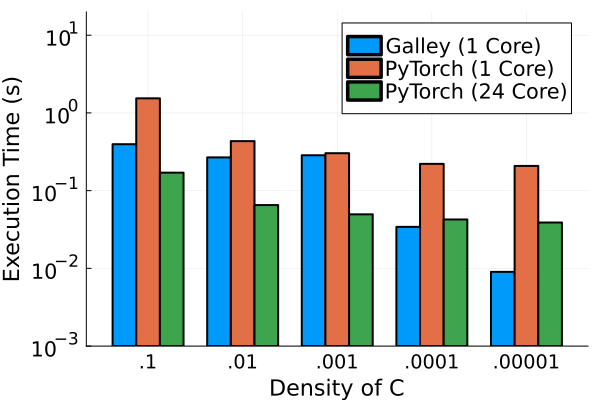}
  \caption{\Ithree{Matrix Chain Multiplication (ABC)}}
  \label{fig:matexp_abc}
\end{subfigure}
\hfill
\begin{subfigure}{.49\textwidth}
  \centering
  \includegraphics[width=\linewidth]{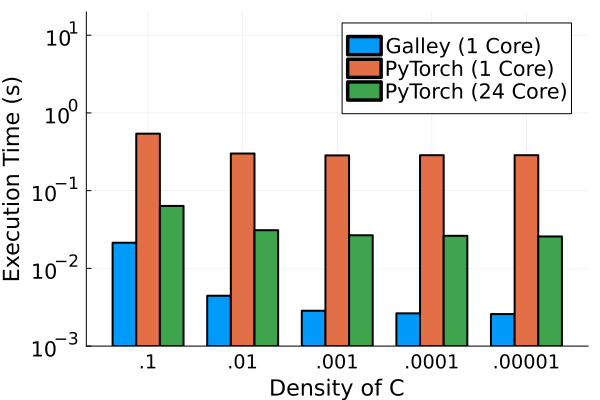}
  \caption{\Ithree{Elementwise Matrix Multiplication (A*B*C)}}
  \label{fig:matexp_elem}
\end{subfigure}
\hfill
\begin{subfigure}{.5\textwidth}
  \centering
  \includegraphics[width=\linewidth]{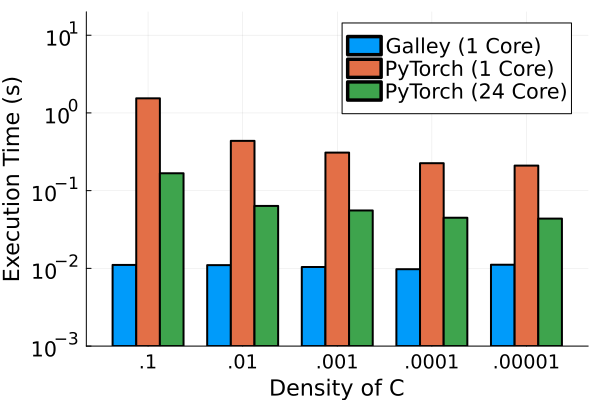}
  \caption{\Ithree{Sum of Matrix Chain (SUM(ABC))}}
  \label{fig:matexp_sum}
\end{subfigure}
\caption{Linear Algebra Kernels}
\label{fig:matexp}
\end{figure*}

\begin{table}[]
\small
\caption{\Ifour{Experimental Dataset Sizes}}
\begin{tabular}{ll}
\toprule
\textbf{Dataset} & \textbf{Size} \\ \midrule
TPCH (SF .25 - SF 5.0)            & .3-6 GB \\
aids             & 11 MB \\ 
human            & 1.5 MB \\ 
yeast            & 1.2 MB \\
dblp             & 21 MB  \\
youtube          & 63 MB  \\
Epinions         & 5.1 MB \\
Kron             & 34 MB  \\
LiveJournal      & .5 GB  \\
Orkut            & 1.7 GB \\
RoadNet          & 41 MB  \\
\bottomrule
\end{tabular}
\label{tbl:datasets}
\end{table}
\section{Experimental Evaluation}
\label{sec:experiments}
In this section, we evaluate the effectiveness of our optimizer on a variety of workloads: (1) ML algorithms over joins, (2) unstructured sparse linear algebra (3) subgraph counting, and (4) breadth-first search. We choose those workloads because they exercise different aspects of our optimizer on real-world use-cases: ML algorithms over joins require careful logical optimizations over programs with mixtures of dense and sparse inputs and non-linear operators; core linear algebra expressions demonstrate the broad utility of Galley; subgraph counting requires both logical and physical optimization of complex sum-product expressions over highly sparse inputs and demonstrates Galley's advantage over a relational engine even for very sparse workloads; breadth-first search requires careful selection of tensor formats over the course of the computation, showing the benefit of physical optimization for even simple computations. Compared to hand-optimized solutions and alternative approaches, Galley is highly computationally efficient while requiring only a concise, declarative input program from the user. Overall, we show that Galley: 
\begin{itemize}[left=0pt]
    \item Performs logical optimizations resulting in $\textbf{1-300}\times$ faster execution for ML algorithms over joins compared to hand-optimized and Pandas implementations and $\textbf{.5-20}\times$ faster runtime when including optimization.
    \item  Optimizes in a mean time of at most $\textbf{0.1}$ seconds for all subgraph counting workloads, with $\textbf{5-20}\times$ faster median execution than DuckDB.
    \item Selects optimal formats for intermediates, outperforming both fully dense and sparse formats for 4/5 graphs in a BFS application.
\end{itemize}

\paragraph{Experiment Setup} These experiments are run on a server with an AMD EPYC 7443P Processor and 256 GB of memory. We implemented Galley in the programming language Julia, and the code is available at \url{https://github.com/kylebd99/Galley.jl}. We used the sparse tensor compiler Finch\footnote{\url{https://github.com/FinchTensor/Finch.jl}} for execution, and all experiments are executed using a single thread. Unless otherwise stated, Galley uses the chain bound described in Sec.~\ref{subsubsec:chain_bound} for sparsity estimation. Experiments for all methods are run five times, and the mean execution time is reported. We perform all experiments on a warm cache, and we separately report the compilation and optimization times.

\subsection{Machine Learning Over Joins}
\label{subsec:ML over Joins}
To explore end-to-end program optimization, we experiment with simple ML algorithms over joins. \Itwo{This represents a typical machine learning use case where the feature matrix is constructed from a variety of tables stored within an enterprise database that are joined together before training. Prior work has shown that co-optimizing these mixed RA/LA problems can yield significant benefits \cite{schleich2016learning,kumar2015learning,chen2016towards,park2022end,grigorov2023p2d}.} \Ione{For this, we consider two join queries over the TPC-H benchmark: a star join and a self join, at scale factor 5 and .25, respectively.} The star join is expressed as follows, where $L,S,P,O,$ and $C$ are tensors representing the line items, suppliers, parts, orders, and customers tables, respectively:
\begin{align*}
    X_{ij} = \sum_{spoc}L_{ispoc}(S_{sj} + P_{pj} + O_{oj} + C_{cj})
\end{align*}
The non-zero values in $S,P,O$ and $C$ are disjoint along the $j$ axis, so the addition in this expression serves to concatenate features from each source, resulting in 139 features after one-hot encoding categorical features. The self-join query compares line items for the same part based on part and supplier features. In this case, the feature data is a 3D tensor because the data points are keyed by pairs of line items:
\begin{align*}
    X_{i_1i_2j} = \sum_{s_1s_2p}L_{i_1s_1p}L_{i_2s_2p}(S_{s_1j} + S_{s_2j} + P_{pj})
\end{align*}
We consider a range of ML algorithms: (1) linear regression inference, (2) logistic regression inference, (3) covariance matrix calculation, and (4) neural network inference. We implement two versions of each of these using the Finch compiler. The dense version uses a dense feature matrix, and the sparse version uses CSR matrix to compress the one-hot encoding. \Ione{We also implemented two standard baselines; 1) using Pandas for the joins and Numpy for the linear algebra 2) using Polars for the joins and PyTorch for the linear algebra. The latter supports parallelism, so we've included parallel results for it as well (marking the parallel bars with P and serial bars with S).}

These algorithms stress the ability of Galley to handle complex operators and combinations of sparse and dense inputs. The definitions of the feature tensors combine pointwise multiplication and addition, and algorithms like logistic regression and neural networks wrap these definitions in non-linear operators (e.g. relu and sigmoid) and aggregates. Further, while the line item tensor is highly sparse, both the feature and parameter tensors are moderately to fully dense. 

\textbf{Execution Time.} Fig.~\ref{fig:ml_over_joins} shows that the execution time of Galley's optimized programs is $.5-300\times$ faster than the sparse Finch implementation. For the regression/neural network problems, this stems from pushing the multiplication with the parameter vector/matrix down to the feature matrices. For the covariance calculation over the self join, Galley fully distributes the multiplication over the addition then aggregates away the sparse $i_1,i_2$ dimensions. This produces small, dense intermediates which can be used to calculate the covariance efficiently.

\textbf{Optimization Time.} Fig. ~\ref{fig:ml_over_joins} also shows that Galley's optimizer has a reasonable overhead in this setting. Concretely, optimization takes $.5 - 5.0$ seconds across all workloads.

\begin{table}[]
\small
\caption{\Ione{Total Subgraph Counting Execution Time (S)}}
\setlength\tabcolsep{2pt}
\begin{tabular}{lcccc}
    \toprule
     \textbf{Workload}  & \textbf{Galley (Greedy)} & \textbf{Galley+DuckDB}  & \textbf{DuckDB} & \textbf{Umbra 1 (24)} \\
     \midrule
     human & .17 (.43) & .156 & .12 & .04 (.02) \\
     aids & 32.1 (29.43) & 43.32 & 78.16 & 7.47 (1.89) \\
     yeast\_lite & 2.96 (3.85) & 8.91 & 1633 & 367.28 (51.56) \\
     dblp\_lite & 32.31 (30.44) & 39.31 & 3294 & 75.51 (18.23) \\
     youtube\_lite & 240.24 (219.47) & 1591.27 & 17203 & 14208 (13866)\\
     \bottomrule
\end{tabular}
\end{table}

\begin{figure}
    \centering
    \begin{subfigure}{.49\textwidth}
    \includegraphics[width=\linewidth]{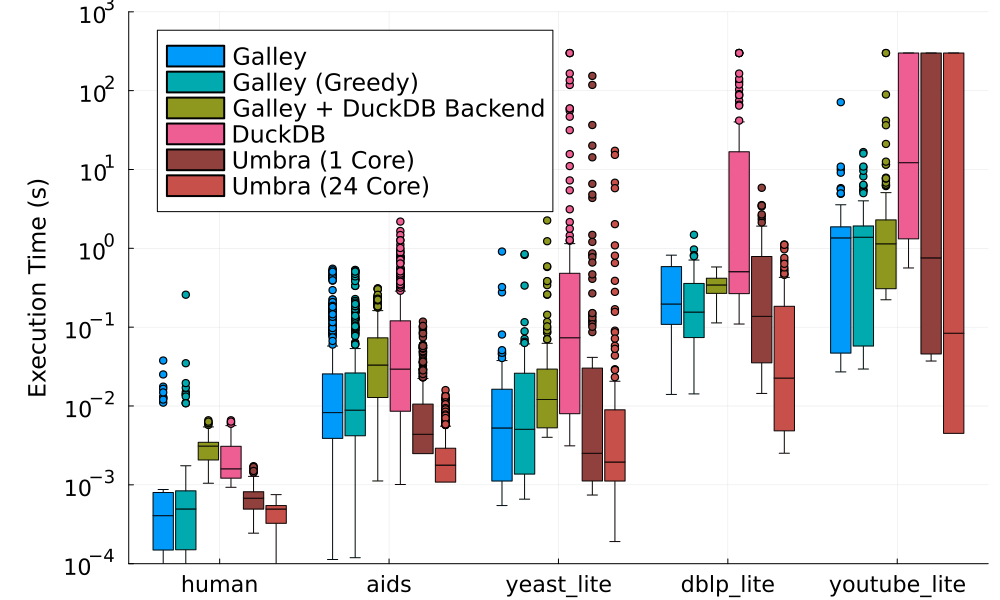}
    \caption{\Ione{Subgraph Counting Execution Time}}
    \label{fig:subgraph_execution}
    \end{subfigure}
    \begin{subfigure}{.49\textwidth}
    \includegraphics[width=\linewidth]{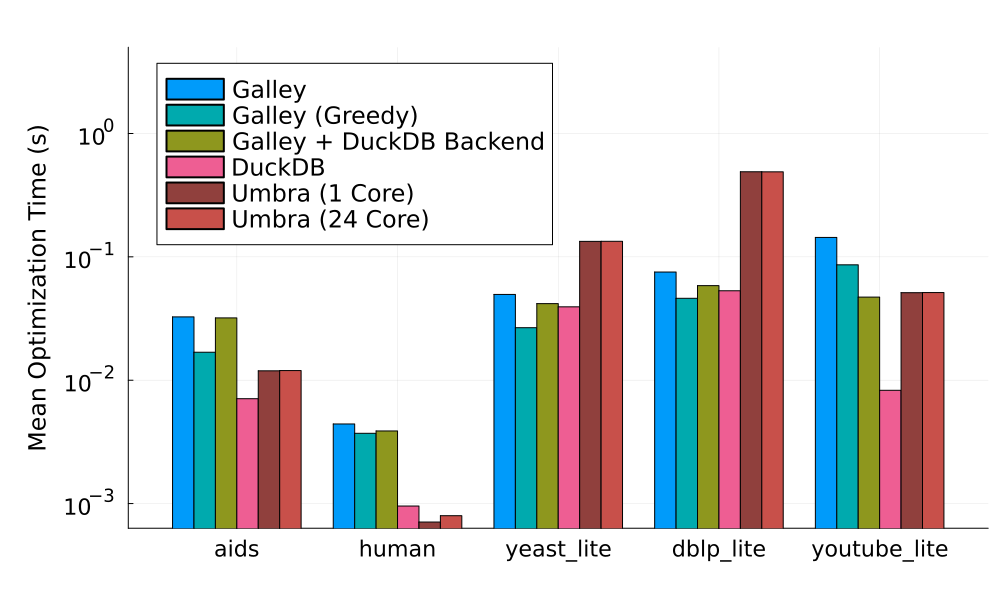}
    \caption{\Ione{Subgraph Counting Optimization Time}}
    \label{fig:subgraph_optimization}
    \end{subfigure} 
    \begin{subfigure}{.49\textwidth}
    \includegraphics[width=\linewidth]{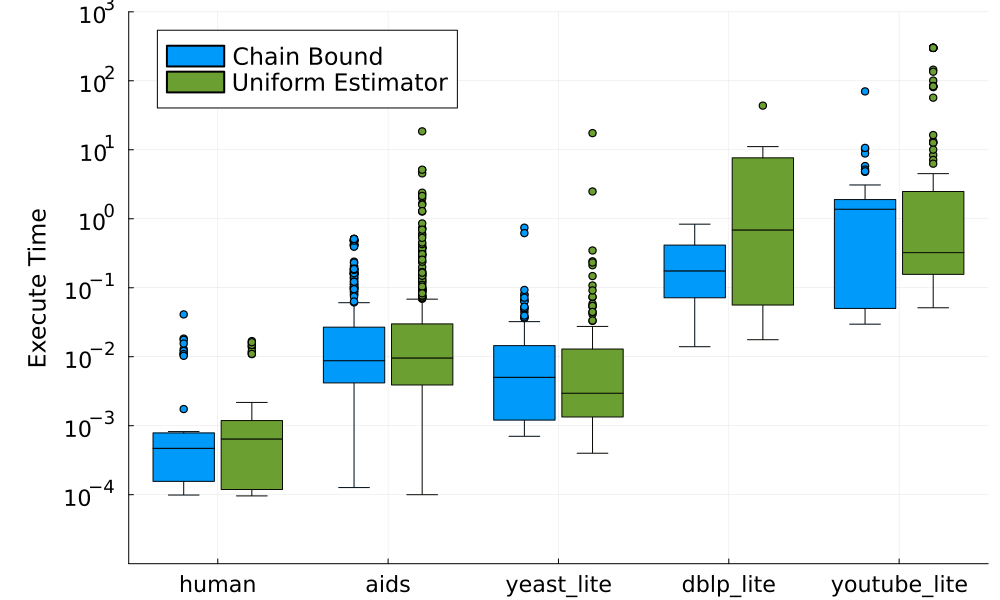}
    \caption{Sparsity Estimator Comparison}
    \label{fig:stats-mb}
    \end{subfigure}  
    \begin{subfigure}{.49\textwidth}
    \includegraphics[width=\linewidth]{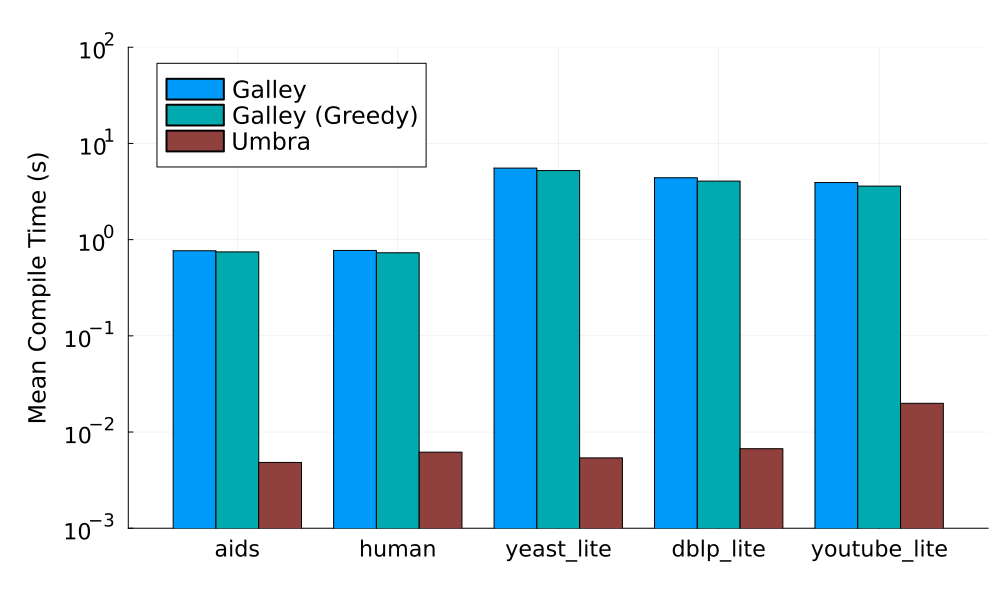}
    \caption{\Ione{Subgraph Counting Compilation Time}}
    \label{fig:subgraph_compilation}
    \end{subfigure}
    \caption{Subgraph Counting Experiments}
\end{figure}
\Ithree{
\subsection{Linear Algebra Kernels}
In Fig.~\ref{fig:matexp}, we show how Galley can provide significant benefits even for simple workloads without structured data. In these experiments, $A$, $B$, and $C$ are $2000\times2000$ uniformly sparse matrices. $A$ and $B$ have a density of $.1$, and the density of $C$ varies on the X axis. In the first experiment, Fig.~\ref{fig:matexp_abc}, Galley improves on PyTorch's execution in two ways: 1) when the matrices are less sparse, it chooses fully dense formats to store then intermediates and outputs 2) when $C$ is heavily sparse, it uses a right-to-left execution strategy for the chain, i.e. $(A(BC))$. In the second experiment, Fig.~\ref{fig:matexp_elem}, Galley is able to fuse the computation when PyTorch is unable to, removing an intermediate materialization. Further, when $C$ is highly sparse, Galley iterates over the non-zeros of $C$ and looks them up in $A$ and $B$, rather than doing a symmetric intersection algorithm. In the third experiment, Fig.~\ref{fig:matexp_sum}, Galley's logical optimizer pushes down the outer summation to $A$ and $C$ first which avoids doing any expensive matrix multiplications. In almost all cases, Galley's optimizations even overcome the benefits of parallelism when PyTorch is provided with 24 cores.
}

\subsection{Subgraph Counting}
\label{subsec:subgraph_counting}
In this section, we stress test Galley's ability to optimize programs with a large number of highly sparse inputs by implementing several sub-graph counting benchmarks. These workloads represent the far end of the complexity and sparsity spectrum for sparse tensor compilers. Suppose you are counting the occurrences of $H(V,E)$ in a data graph $G$ with adjacency matrix $M$; we can represent the count as,
\begin{align*}
    c = \sum_{v_i\in V}\prod_{(v_i,v_j)\in E}M_{v_iv_j}
\end{align*}
We add sparse binary vector factors for each labeled vertex. We use subgraph workloads from the G-Care benchmark and the paper "In-Memory Subgraph: an In-Depth Study"~\cite{DBLP:conf/sigmod/Sun020,DBLP:conf/sigmod/ParkKBKHH20}. We restrict them to query graphs with up to 8 vertices and 24 edges. \Ione{Because this is a relational workload, we compare it with DuckDB and Umbra, two state-of-the-art modern OLAP databases~\cite{DBLP:conf/sigmod/RaasveldtM19,DBLP:conf/cidr/NeumannF20}. The latter is known to be one of the fastest databases for complex joins and aggregations, and we include both serial and parallel execution~\cite{ClickBench}. We did not include these systems in our other experiments due to the difficulty of framing the problems in SQL and because prior work has already demonstrated their challenges on pure LA workloads~\cite{DBLP:journals/corr/abs-2312-17355}.} To separately discern the impact of logical vs physical optimization and our use of Finch, we provide a version of Galley that executes each logical query with a SQL query run on DuckDB. We also provide results for the greedy logical optimizer.

\textbf{Logical Optimization.} Fig.~\ref{fig:subgraph_execution} shows the execution time for each method and benchmark. The comparison between `DuckDB` and `Galley + DuckDB Backend` demonstrates the benefits of Galley's logical optimizer. Galley's logical optimizer breaks down the program into a series of aggregations which minimizes the necessary computation and materialization. This has the largest impact on graphs with high skew like the social network graphs, `dblp\_lite` and `youtube\_lite`. In these cases, pushing down aggregates avoids very large intermediate results. DuckDB hits the 300 second timeout on 56 out of 120 queries in the youtube\_lite benchmark, as does Umbra on 46 queries. In contrast, Galley never times out across all workloads.

\textbf{Physical Optimization.} The impact of Galley's physical optimizer can be seen by comparing `Galley` with `Galley + DuckDB`. Galley's median execution is up to 8x faster than DuckDB even with the same logical plan. This shows that Galley is selecting efficient loop orders and formats, effectively leveraging STCs.

\textbf{Optimization Time.} Fig.~\ref{fig:subgraph_optimization} shows the mean optimization time for each method on each workload. Galley has a mean optimization time of less than $.15$ seconds across all workloads, faster than Umbra's optimizer for 2 workloads.

\textbf{Compilation Time.} Because it performs compilation at runtime, Galley incurs a compilation overhead when it invokes an STC kernel. These kernels are cached by Finch, reducing this cost when workloads repeatedly use similar kernels. We show the mean compilation time for each subgraph workload in Fig. \ref{fig:subgraph_compilation}. On the simpler workloads, which often reuse kernels, this cost is lower. More complex workloads reuse kernels less, significantly increasing compilation time.

\Ifour{Comparing Figures 8 and 9, Galley’s optimization overhead is minimal (generally less than 1\%) compared to the compilation overhead. Reducing this requires performance improvements to the underlying compiler which are out of scope for this work. Fortunately, the Finch project is working to improve this in two ways (1) by caching compilation to disk and (2) by migrating to the MLIR compiler infrastructure. As these improvements are made, Galley will immediately reap the benefits.}

\textbf{Sparsity Estimation.} Finally, in Fig.~\ref{fig:stats-mb}, we use the sub-graph matching workloads to compare sparsity estimators and their effect on performance. Across all workloads, we see that the chain bound has significantly better tail performance. This is because it encourages more conservative query plans which better handle correlated and skewed queries/datasets.

\begin{figure}
    \centering
    \includegraphics[width=.6\linewidth]{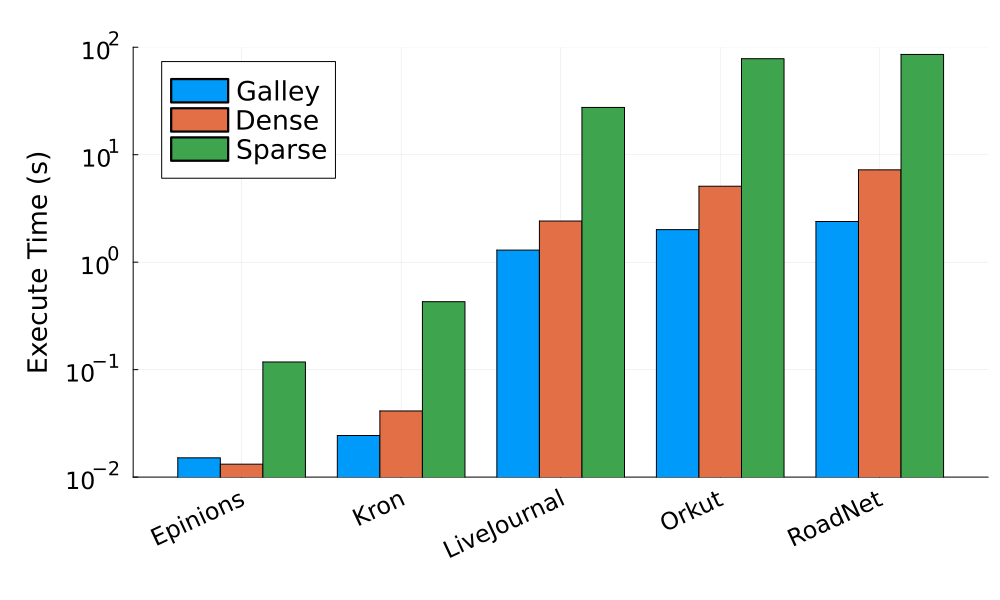}
    \caption{BFS Execution Time}
    \label{fig:bfs}
\end{figure}

\subsection{Breadth-First Search}
To demonstrate the importance of format selection, we implement a breadth-first search algorithm using Galley and hand-coded Finch implementations. Both systems receive a single iteration at a time, and the total execution time is reported. The core computation is a masked sparse matrix times sparse vector to compute the new frontier vector. The main decision is the visited and frontier vectors' formats. The former's sparsity grows monotonically over iterations, while the latter peaks in the middle iterations. We provide two implementations of Finch, using either a sparse or a dense vector for both. Fig.~\ref{fig:bfs} shows that Galley's mixture of sparse and dense formats is significantly fastest for 4 of the 5 graphs and is competitive for all graphs. For 4/5 graphs, the total optimization time (not depicted in the figure) is less than $.25$ seconds. This experiment demonstrates the utility of sparsity-aware format selection, and future work should consider ways to amortize optimization time for iterative workloads.

\section{Related Work}
Galley differs from other work on cost-based optimization for tensor processing due to its targeting of STCs and its expressive input language. SystemDS, formerly SystemML, focuses on end-to-end ML over matrices and tabular data \cite{DBLP:journals/pvldb/BoehmDEEMPRRSST16, DBLP:conf/cidr/BoehmADGIKLPR20, DBLP:journals/pacmmod/Baunsgaard023}; it takes as input linear algebra (LA) programs and targets a combination of LA libraries and distributed computing via Spark. Later work, SPORES, extended its logical optimizer to leverage relational algebra when optimizing sum-product expressions\cite{DBLP:journals/pvldb/WangHSHL20}; their core insight was that LA rewrites, which always match and produce 0-2D expressions, are not sufficient and that optimal rewrites must pass through higher order intermediate expressions. Other related work translated sum-product expressions to SQL to leverage highly efficient database execution engines \cite{blacher2023efficient}. These systems can perform well for highly sparse inputs but struggle on mixed dense-sparse workloads. Tensor relational algebra proposes a relational layer on top of dense tensor algebra that provides a strong foundation for automatically optimizing distributed dense tensor computations \cite{DBLP:journals/pvldb/YuanJZTBJ21,bourgeois2024eindecomp}. The compiler community has made attempts to automatically optimize sparse tensor sum-product kernels based on asymptotic performance analyses\cite{ahrens_autoscheduling_2022,DBLP:journals/corr/abs-2311-09549}. These systems each target a different execution context and focus on different aspects of optimization. Galley expands on this line of work by targeting a new execution engine, proposing novel optimization techniques, and handling a wider range of tensor programs.

\section{Limitations} 

We are excited to enrich Galley with new optimizations in the future. Currently, Galley lacks support for complex loop structures (e.g.,  a single outer FOR loop that wraps multiple inner FOR loops), higher order functions (e.g. matrix inversion) or parallelism. However, we believe that these areas could benefit from cost-based optimization. Similarly, Galley does not consider hard memory constraints during optimization, but our use of cardinality bound methods provides an avenue for addressing this in future work.

\bibliographystyle{ACM-Reference-Format}
\bibliography{main}

\received{October 2024}
\received[revised]{January 2025}
\received[accepted]{February 2025}

\appendix

\end{document}